# Slicing Event Traces of Large Software Systems

Raymond Smith
Lucent Technologies
Naperville, Illinois, USA
rds1@lucent.com

Bogdan Korel
Computer Science Department
Illinois Institute of Technology
Chicago, Illinois USA
korel@iit.edu

**Abstract**
Debugging of large software systems consisting of many processes accessing shared resources is a very difficult task. Many commercial systems record essential events during system execution for post-mortem analysis. However, the event traces of large and long-running systems can be quite voluminous. Analysis of such event traces to identify sources of incorrect behavior can be very tedious, error-prone, and inefficient. In this paper, we propose a novel technique of slicing event traces as a means of reducing the number of events for analysis. This technique identifies events that may have influenced observed incorrect system behavior. In order to recognize influencing events several types of dependencies between events are identified. These dependencies are determined automatically from an event trace. In order to improve the precision of slicing we propose to use additional dependencies, referred to as cause-effect dependencies, which can further reduce the size of sliced event traces. Our initial experience has shown that this slicing technique can significantly reduce the size of event traces for analysis.

## 1   Introduction

Large, long-running computer systems used for such applications as real-time operations, process control, or online transaction processing may consist of millions of lines of code in many separate processes. Many different system components may be accessed and used by any number of these independent processes and the coordination and analysis of the interactions of these components is often necessary for debugging and other maintenance activities. Such large software systems record events as they occur during execution resulting in a dynamic record of system behavior that can be used for debugging and problem analysis. An event trace is a record of events associated with operations that are performed upon various entities of interest such as physical devices (e.g., networks, circuit packs, peripheral units) or software resources (e.g., buffers in memory, individual processes, sets of processes). Event traces are used for analysis in many systems [And98, Buk95, Coo98]. Often, the event trace may be the only artifact available for post-mortem analysis and debugging of the system, but the event traces for large, complex, long-running systems can be quite voluminous, particularly if there are a large number of independent processes and shared resources. Manual analysis involves sorting through a long list of disparate events to find those that are related to the problem under study. The experience of one of the authors of this paper confirms that manual analysis can be time consuming, error prone and tedious. While several tools exist for detecting anomalies and constraint violations in the event trace, [Bal96, Len00], these tools do not reduce the size of the event trace for analysis. Frequently such tools do not identify any constraint violations even though an incorrect behavior was observed. Since during a debugging session, developers have to analyze long event traces, they are very much interested in reducing the number of events to be analyzed.

One well-known reduction technique is program slicing. A program slice consists of all statements in a program that may affect the value of a variable at some point. There are two types of program slices: static slices and dynamic slices. A static slice [Wei82] preserves the program's behavior with respect to a selected variable for all program inputs, and it is used to identify these parts of the program that potentially contribute to the computation of the selected variable. On the other hand, a dynamic slice [Kor88] preserves the program's behavior with respect to a variable for particular program input, and it is used to identify these parts of the program that contribute to the computation of the selected variable for a given program execution (program input). Dynamic slices are frequently much smaller than static slices. Several methods of computation of static slices have been proposed in the literature [Gal91, Hor90, Wei82]. Similarly, several algorithms to compute dynamic slices have been proposed [Agr90, Kor88, Kor97a]. Originally, program slicing has been proposed to guide programmers during program debugging [Agr93, Lyl86], but program slicing can also be used in the process of understanding of programs during software maintenance and testing [Gal91, Kor97, Whi92].

Traditional static and dynamic slicing techniques are used to reduce the program (source code) size. Dynamic slicing can be also used to reduce the size of an execution trace. However, the existing slicing techniques cannot be used to reduce event traces. In this paper, we propose a novel slicing technique on event traces as a means of reducing the number of events for analysis. This technique identifies events that may have influenced observed incorrect system behavior, constructing a sliced event trace consisting of such influencing events. In order to recognize influencing events several different types of dependencies between events are identified. These dependencies can be determined fully automatically from an event trace. In order to improve the precision of slicing we propose to provide additional cause-effect dependencies that can further reduce the size of a sliced event trace. Event trace reduction may make debugging significantly easier due to the smaller size of event traces required for analysis. Our initial experience has shown that this slicing technique can significantly reduce the size of event traces for analysis.

In Section 2, a system model and a structure of an event trace is provided. In Section 3 dependencies between events are defined, and in Section 4 the slicing algorithm that uses these dependencies is described. Section 5 introduces cause-effect dependencies and their application for slicing of event traces.

## 2    System Model and Event Traces

A typical system consists of a set of *active resources* $R_A$ and *passive resources* $R_P$. An active resource is a system component such as a process that can perform some action or operation on another resource or on itself. A passive resource is a system component that is acted upon such as a file or a peripheral unit, but that does not initiate any action or operation itself. Each resource is characterized by a set of states and a set of operations that can be performed on it which may cause the state of that resource to change. As a result, each resource is modeled by a state transition diagram that captures state changes in the resource based on performed operations. State transition diagrams may be obtained from requirements or design documentation and be constructed for individual resources or for a class of similar resources, e.g., all files, all processes. Figure 4a and 4b show state transition diagrams for a process resource and for a file resource, respectively. Formally, a resource is modeled as $R = <S, O, T>$, where $S = \{s_1, s_2, ..., s_m\}$ is a set of states that the resource can take on, $O = \{o_1, o_2, ..., o_n\}$ is a set of operations that can be performed on the resource, and $T = \{(o_1, s_1, d_1), (o_2, s_2, d_2), ..., (o_l, s_l, d_l)\}$ is a set of transitions (triples) where $s_i, d_i \in S$, $o_i \in O$, and $(o_i, s_i, d_i)$ indicates that operation $o_i$ causes the state of the resource to change from $s_i$ to $d_i$.

A system model consists of a set of resources and relationships between them that model interactions between resources. A relationship between resources indicates that one resource *may interact* with another and this interaction may cause the second resource to change its state. For example, a process can issue a *Print* operation on a printer to cause its state to change from *Idle* to *Printing*, or a process initiates some interaction with another process such as sending an abort signal that may cause a state change in this process from *Running* to *Unavailable*.

Attempts have been made to standardize event traces [Cou92, Ree91], particularly in the area of performance analysis and parallel system design, but no real standard has been established. In this paper, we assume that an event trace has the following format that, probably, represents a typical format used in many commercial systems. By an event trace $T$ we mean an ordered sequence of events, $T = <E_1, E_2, E_3, ..., E_m>$ where $E_i$ is an event. For purposes of our analysis, we assume that for each event $E_i$ the following information is recorded: $E_i = <P, O, R, S_O, S_N>$ where, $P$ is an identifier of the process executing an operation $O$, $O$ is the operation identifier, $R$ is the passive or active resource identifier, $S_O$ is the old state of resource $R$ before the operation was performed, and $S_N$ is the new state of the resource $R$ after the operation was performed. An event $E_i$ is considered to have occurred whenever a process $P$ executes an operation $O$ on resource $R$ causing the state of R to change from $S_O$ to $S_N$ after completion of operation $O$. For example, the following event may be recorded: $<P2, Open, FileE, Closed, Open>$, which shows process $P2$ performed an *Open* operation on the resource *FileE*, which resulted in a change of state for the resource *FileE* from *Closed* to *Open*. The following notation is used in the paper to refer to individual components of an event $E_i$: $P(E_i)$, $O(E_i)$, $R(E_i)$, $S_O(E_i)$, and $S_N(E_i)$ denote a process identifier, operation identifier, resource identifier, old state, and new state identifier of event $E_i$, respectively. Figure 1 shows an event trace for a particular execution run. Notice that some operations do not cause a state change to occur and so the old and new state are the same.

## 3    Dependencies Between Events

Dynamic slicing [Kor88] is a reduction technique that uses dependencies between source code statements as a way of determining whether to include or exclude these statements from an execution trace. Statements that belong to a sliced execution trace are a part of a dynamic slice. To use this technique on event traces, it is necessary to identify different dependencies between events. These dependencies may be then used to determine whether or not to include events as part of a sliced event trace. This approach may narrow the size of the event trace and assist in determining what caused the event with an incorrect behavior. Three types of dependencies are identified that are easily discernible from the given trace.

### 3.1 Change-Of-State dependency

A Change-Of-State (COS) dependency captures a situation when in one event a certain resource state is observed and we would like to identify the last event where the change of the observed resource state took place. Assume there are two events $E_i$ and $E_j$, $i < j$, in an event trace. A COS dependency exists between $E_i$ and $E_j$ when:

1) $R(E_i) = R(E_j)$ and $S_n(E_i) = S_o(E_j)$,
2) $S_o(E_i) \neq S_n(E_i)$, and
3) For all $k$, $i<k<j$, if $(R(E_i) = R(E_k))$ then $(S_o(E_k) = S_n(E_i) = S_n(E_k))$.

Condition 1 ensures that the same resource state appears in both events. Condition 2 ensures that a change of the resource state took place in $E_i$, and condition 3 ensures that the resource state has not changed between these events. In the event trace of Figure 1, a COS dependency exists between $E_5$ and $E_{32}$, and $E_1$ and $E_{33}$.

### 3.2 Last-Resource-Use dependency

A change of state occurred in a process (active resource). We want to identify recently used resources by the process. This can be captured by the Last-Resource-Use (LRU) dependency. Assume there are two events $E_i$ and $E_j$, $i < j$, in an event trace. A LRU dependency exists between events $E_i$ and $E_j$ when:

1) $R(E_j)$ is an active resource, and $S_O(E_j) \neq S_N(E_j)$,
2) $P(E_i) = R(E_j)$, and
3) For any $k$, $i<k<j$, if $P(E_k) = R(E_j)$ then $R(E_i) \neq R(E_k)$.

Condition 1 ensures the resource in $E_j$ is an active resource and a state change has occurred in $E_j$. Condition 2 requires that the active resource in $E_j$ is the process performing an operation in $E_i$. Condition 3 ensures that no other event exists between $E_i$ and $E_j$ where the active resource in $E_i$ uses the passive resource $R(E_i)$. For example, in the event trace of Figure 1, an LRU dependency exists between $E_{28}$ and $E_{33}$.

### 3.3 Last-Shared-Resource-Use dependency

A change of state occurred in a process (active resource) that has used a resource recently. We want to identify whether this resource is shared with another process and identify the last use of this resource by another process. This can be captured by the Last-Shared-Resource-Use (LSRU) dependency. Assume there are two events $E_i$ and $E_j$, $i < j$, in an event trace. A LSRU dependency exists between events $E_i$ and $E_j$ when:

1) There exists event $E_m$, $m<j$, such that a LRU dependency exists between $E_m$ and $E_j$,
2) $P(E_m) \neq P(E_i)$, and $R(E_m) = R(E_i)$,
3) a) if $i<m$: for all $k$ (except $k=m$), $i<k<j$, $R(E_i) \neq R(E_k)$, or
   b) if $i>m$: for all $k$, $i<k<j$, $R(E_i) \neq R(E_k)$.

Condition 1 ensures that for a LSRU dependency to exist between $E_i$ and $E_j$, a LRU dependency must exist between $E_j$ and another event $E_m$. Conditions 2 ensures that the processes in $E_m$ and $E_i$ are different but they use the same shared resource $R(E_i)$. Condition 3a addresses a situation when $E_i$ occurs before $E_m$ and ensures that no other process between $E_i$ and $E_j$ uses the shared resource except for $E_m$. Condition 3b addresses a situation when $E_i$ occurs after $E_m$ and ensures that no other process between $E_i$ and $E_j$ uses the shared resource. For example, in Figure 1, a Last-Shared-Resource-Use dependency exists between $E_{24}$ and $E_{33}$.

## 4 Slicing Event Traces

A typical debugging session requiring analysis of an event trace takes place when an incorrect behavior is observed and an event(s) that may have caused this behavior needs to be identified. All three dependencies described in the previous section are used by a slicing algorithm to identify events influencing an observed behavior. The slicing algorithm starts with an initial event $E_s$ in which an incorrect behavior is observed. All events in the event trace are set as unmarked and not visited. In the first step, events that have a COS, LRU or LSRU dependency on the initial event $E_s$ are identified and marked. In the next step, a marked but not visited event is selected. Events that have influenced the selected event by COS, LRU, or LSRU dependencies are identified and marked (if not marked before). The selected event is set as visited. In the next step a marked but not visited event is selected and new events with dependencies on the selected event are identified and marked. This process of identifying and marking events is repeated until all marked events are visited. All events that are unmarked do not affect the starting event and are removed from the event trace. The resulting event sub-trace consists of events that may have potentially influenced the initial event $E_s$. Suppose that in the event trace of Figure 1 the event $E_{37}$, indicating that *P1* has changed state from *Blocked* to *Running*, requires investigation. It is necessary to determine as to what caused this change of state by determining which events in the trace may have caused *P1* to be *Blocked*. Figure 2 shows the sliced sub-trace of the event trace of Figure 1 after applying the slicing algorithm. Figure 3 shows the dependencies that were identified and used by the slicing algorithm, resulting is a size reduction of the event trace from 37 events to 15.

| No. | Proc. | Oper. | Rsrc. | Old State | New State | No. | Proc. | Oper. | Rsrc. | Old State | New State |
|-----|-------|-------|-------|-----------|-----------|-----|-------|-------|-------|-----------|-----------|
| 1 | P3 | Start | P1 | Unav. | Running | 20 | P2 | Write | FileE | Locked | Locked |
| 2 | P2 | Start | P2 | Unav. | Running | 21 | P2 | Write | FileC | Locked | Locked |
| 3 | P2 | Mount | FSys1 | Unav. | Mounted | 22 | P1 | Read | FileA | Open | Open |
| 4 | P2 | Mount | FSys2 | Unav. | Mounted | 23 | P1 | Read | FileB | Open | Open |
| 5 | P1 | Open | FileA | Closed | Open | 24 | P2 | Write | FileC | Locked | Locked |
| 6 | P1 | Open | FileB | Closed | Open | 25 | P1 | Read | FileA | Open | Open |
| 7 | P2 | Open | FileC | Closed | Open | 26 | P2 | Unlock | FileE | Locked | Open |
| 8 | P2 | Open | FileD | Closed | Open | 27 | P1 | Write | FileC | Locked | Locked |
| 9 | P2 | Open | FileE | Closed | Open | 28 | P1 | Lock | FileC | Locked | Locked |
| 10 | P1 | Read | FileA | Open | Open | 29 | P1 | Read | FileB | Open | Open |
| 11 | P2 | Read | FileC | Open | Open | 30 | P1 | Lock | FileB | Open | Locked |
| 12 | P1 | Read | FileB | Open | Open | 31 | P1 | Write | FileB | Locked | Locked |
| 13 | P2 | Lock | FileC | Open | Locked | 32 | P1 | Read | FileA | Open | Open |
| 14 | P1 | Read | FileA | Open | Open | 33 | P1 | Wait | P1 | Running | Blocked |
| 15 | P1 | Lock | FileB | Open | Locked | 34 | P2 | Write | FileD | Locked | Locked |
| 16 | P1 | Write | FileB | Locked | Locked | 35 | P2 | Write | FileC | Locked | Locked |
| 17 | P1 | Unlock | FileB | Locked | Open | 36 | P2 | Unlock | FileC | Locked | Open |
| 18 | P2 | Lock | FileE | Open | Locked | 37 | P1 | Signal | P1 | Blocked | Running |
| 19 | P2 | Lock | FileD | Open | Locked | | | | | | |

**Figure 1** Event Trace.

## 5 Cause-Effect Dependencies

Two types of dependencies LRU and LSRU are used to identify recently used resources because their usage may be a cause of incorrect behavior. However, a large number of events may be identified by these dependencies, many of which are not responsible for the incorrect behavior. This may lead to inclusion of too many events in a sliced event trace. There is a need to distinguish suspicious behavior from normal behavior. Frequently, only a certain usage of a resource may lead to an incorrect behavior of a process, e.g., locking a file may block a process, however reading a file may not block the process. We propose to incorporate additional information, referred to as *cause-effect dependencies*, that can be used to identify suspicious resource usage, and as a result, reduce the size of the event trace. In general, a cause-effect dependency between resources indicates that a certain usage of one resource may be responsible for a certain behavior of the other. We distinguish two types of cause-effect dependencies: *static* and *dynamic*. Static cause-effect dependencies are defined between elements of state transition diagrams whereas dynamic dependencies are defined between events in the event trace.

### 5.1 Static cause-effect dependencies

A cause-effect dependency may only exist between resources for which a *may interact* relationship exists. Identifying static cause-effect dependencies between elements of the state transition diagrams of two resources may be treated as a refinement of the *may interact* relationship between these resources. Static cause-effect dependencies have to be identified by developers. Four types of cause-effect dependencies can be identified between elements of state transition diagrams: state-to-state, state-to-transition, transition-to-state and transition-to-transition cause-effect dependencies. In this paper, we present only the transition-to-transition cause-effect dependency, but remaining dependencies can be similarly defined.

A transition-to-transition cause-effect dependency between resources indicates that a usage of one resource resulting in a specific transition may be responsible for a specific change of state of another resource. Assume $(o_i, s_i, d_i)$ is a transition in the state transition diagram of resource $R$. Additionally, assume $(o_k, s_k, d_k)$ is a transition in the state transition diagram of process $P$. A transition-to-transition cause-effect dependency between $(o_i, s_i, d_i)$ and $(o_k, s_k, d_k)$ indicates that a usage of resource $R$ resulting in a transition $(o_i, s_i, d_i)$ may be the reason transition $(o_k, s_k, d_k)$ occurred in process $P$. For example, in the state diagram of Figure 4c, a *Lock* operation on a file resulting in a transition from *Locked* to *Locked* may cause a transition from *Running* to *Blocked* to occur in a process. Therefore, a cause-effect dependency may be introduced between these transitions. Two of these transition-to-transition cause-effect dependencies are shown graphically in Figure 4c as dotted arrows. Similarly, state-to-state, state-to-transition and transition-to-state cause-effect dependencies may be introduced.

| No. | Proc. | Oper. | Rsrc. | Old State | New State |
|-----|-------|--------|-------|-----------|-----------|
| 1   | P2    | Start  | P1    | Unav.     | Running   |
| 5   | P1    | Open   | FileA | Closed    | Open      |
| 6   | P1    | Open   | FileB | Closed    | Open      |
| 7   | P2    | Open   | FileC | Closed    | Open      |
| 13  | P2    | Lock   | FileC | Open      | Locked    |
| 15  | P1    | Lock   | FileB | Open      | Locked    |
| 17  | P1    | Unlock | FileB | Locked    | Open      |
| 24  | P2    | Write  | FileC | Locked    | Locked    |
| 28  | P1    | Lock   | FileC | Locked    | Locked    |
| 30  | P1    | Lock   | FileB | Open      | Locked    |
| 31  | P1    | Write  | FileB | Locked    | Locked    |
| 32  | P1    | Read   | FileA | Open      | Open      |
| 33  | P1    | Wait   | P1    | Running   | Blocked   |
| 36  | P2    | Unlock | FileC | Locked    | Open      |
| 37  | P1    | Signal | P1    | Blocked   | Running   |

**Figure 2** Sliced event trace

| Dependencies used | Dependent events |
|-------------------|------------------|
| Change-Of-State | (33,37), (30,31), (17,30), (15,17), (13,36), (7,13), (6,15), (5,32), (1,33) |
| Last-Resource-Use | (32,37), (31,37), (28,37) |
| Last-Shared-Resource-Use | (36,37), (24,33) |

**Note:** Each dependent event pair ($E_i$, $E_j$) indicates that $E_j$ depends upon $E_i$.

**Figure 3** Dependencies used to compute the sliced event trace of Figure 2.

### 5.2 Dynamic cause-effect dependencies

The static cause-effect dependencies between elements of state transition diagrams of resources may be used to automatically identify dynamic cause-effect dependencies between events and achieve further reduction of the sliced event trace. A dynamic cause-effect dependency between events indicates that a certain usage of a resource may likely be responsible for a certain behavior of an active resource because a static cause-effect dependency exists between state transition diagram elements of these resources. Dynamic cause-effect dependencies represent a subset of LRU and LSRU dependencies that most likely are responsible for incorrect system behavior. Formally, a dynamic cause-effect dependency is defined as follows: assume there are two events $E_i$ and $E_j$, $i<j$. A dynamic cause-effect dependency exists between $E_i$ and $E_j$ $i < j$, if, (1) there exists an LRU or LSRU dependency between $E_i$ and $E_j$, and (2) there exists a static cause-effect dependency between elements of $E_i$ and $E_j$.

Consider event $E_{37}$ in the event trace of Figure 1. For this event there exists the following LRU and LSRU dependencies: ($E_{28}$, $E_{37}$), ($E_{31}$, $E_{37}$), ($E_{32}$, $E_{37}$), and ($E_{36}$, $E_{37}$). When the static, transition-to-transition cause-effect dependencies of Figure 4c are used, only one dynamic cause-effect dependency is identified: ($E_{36}$, $E_{37}$). This is included because a static cause-effect dependency exists between the transitions in $E_{36}$ and $E_{37}$. No such dependency exists between any of the other events with an LRU or LSRU dependency on $E_{37}$. The use of dynamic cause-effect dependencies can result in significant reduction of the sliced event trace. Using the transition-to-transition cause-effect dependencies of Figure 4c on event $E_{37}$ results in an event sub-trace containing only 6 events, substantially less than the 15 events for the same starting point as shown in Figure 5. Figure 6 lists all dynamic cause-effect dependencies that were used to determine the sliced event trace shown in Figure 5.

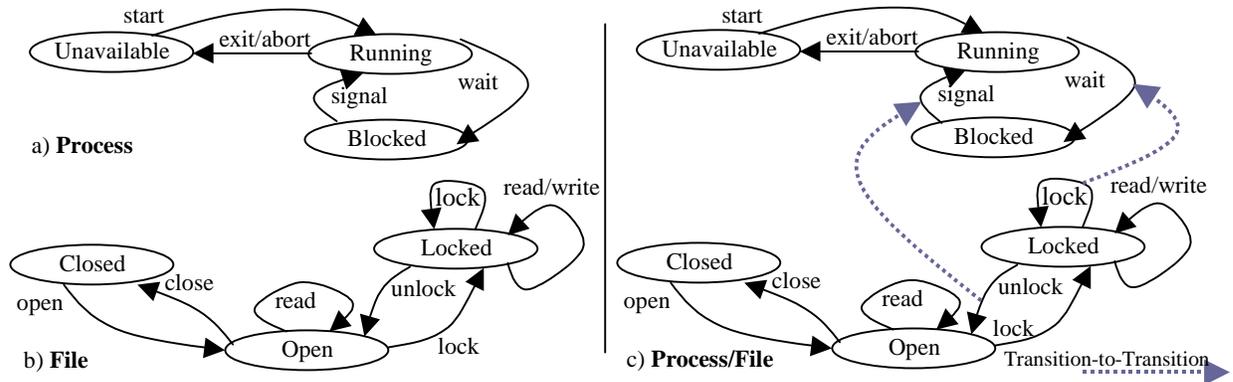

**Figure 4** State transition diagrams for resources with Cause-Effect dependencies.

| No. | Proc. | Oper. | Rsrc. | Old State | New State |
|---|---|---|---|---|---|
| 1 | P2 | Start | P1 | Unav. | Running |
| 7 | P2 | Open | FileC | Closed | Open |
| 13 | P2 | Lock | FileC | Open | Locked |
| 33 | P1 | Wait | P1 | Running | Blocked |
| 36 | P2 | Unlock | FileC | Locked | Open |
| 37 | P1 | Signal | P1 | Blocked | Running |

**Figure 5** Sliced event trace using cause-effect dependencies of Figure 4c.

| Dependencies used | Dependent events |
|---|---|
| Cause-effect (transition-to-transition) | (36,37) |
| Change-Of-State | (33,37), (13,36), (7,13), (1,33) |

**Figure 6** Dependencies used to compute the sliced event trace of Figure 5.

## 6 Conclusions

In this paper we presented a novel slicing technique to reduce the size of event traces for debugging of large software systems. Several types of dependencies between events have been identified that are used by a slicing algorithm. Additional cause-effect dependencies were introduced that can further reduce the size of the sliced event trace. Cause-effect dependencies are provided by developers, but in many cases identification of such dependencies may not require major effort and may be derived from the existing system documentation. Cause-effect dependencies may frequently be introduced only for a class of resources, not for individual resources. Identifying cause-effect dependencies among resources may be approached as an incremental process with successive debugging sessions identifying more dependencies to add to a growing base of information. This effort can pay off in smaller event traces for analysis that may significantly reduce debugging effort. We are in the process of developing a tool that will automatically perform slicing of event traces. We plan to perform experiments to determine the effectiveness of this slicing technique in the reduction of event traces of commercial systems. Our approach is currently limited to slicing of single event traces. One of the extensions is to develop slicing techniques for multiple event traces created by multiprocessing systems. We also plan to investigate the trade-offs between the amount of information recorded and the precision of slicing.